\documentclass[11pt,a4paper]{article}

 \newcommand{\be}{\begin{equation}}
 \newcommand{\ee}{\end{equation}}
 \newcommand{\bl}{\begin{equation}\begin{array}{ll}}
 \newcommand{\el}{\end{array}\end{equation}}
 \newcommand{\bll}{\begin{equation}\begin{array}{lll}}
 \newcommand{\bdm}{\begin{displaymath}}
 \newcommand{\edm}{\end{displaymath}}
 \def\bea{\begin{eqnarray}}
 \def\eea{\end{eqnarray}}
 \def\barr{\begin{array}}
 \def\earr{\end{array}}
\def\p{\partial}

 \def\de{\delta}

\def\half{\frac{1}{2}}

\def\2third{\frac{2}{3}}
\def\4third{\frac{4}{3}}
\def\3quart{\frac{3}{4}}

\def\sixth{\frac{1}{6}}

\def\lim{\rightarrow}

\def\cL{{\cal L}}

\def\dx{\dot{x}}

\def\ddx{\ddot{x}}

\def\drho{\dot{\rho}}

\def\dsig{\dot{\sigma}}

\def\dal{\dot{\alpha}}
\def\dbe{\dot{\beta}}

\def\dA{\dot{A}}

\def\hba{\hat\textbf{a}}

\def\hga{\hat{\gamma}}

\def\ha{\hat{a}}

\def\Lag{{\cal L}}

\def\ta{\tilde{a}}
\def\ts{\tilde{s}}
\def\tm{\tilde{m}}

\begin{document}
\raggedbottom

\title{{\bf An old Einstein - Eddington generalized gravity \\
 and modern ideas on branes and cosmology.}}

\author{A.T.~Filippov \thanks{Alexandre.Filippov@jinr.ru}~ \\
{\small \it {$^+$ Joint Institute for Nuclear Research, Dubna, Moscow
Region RU-141980} }}

\maketitle

\begin{abstract}
 We briefly discuss new models of an `affine' theory of gravity in
 multidimensional space-times  with \emph{symmetric connections}.
 We use and generalize Einstein's proposal
 to specify the space-time geometry by use of the Hamilton
 principle to determine the connection coefficients
 from a \emph{geometric Lagrangian} that is an arbitrary
 function of  the  generalized Ricci curvature
 tensor and of other  fundamental  tensors.
  Such a theory supplements the standard Einstein gravity
 with dark energy (the cosmological constant, in the first
 approximation), a neutral massive (or tachyonic) vector field
 (\emph{vecton}), and massive (or tachyonic)
 \emph{scalar fields}.
 These fields couple only to
 gravity  and can generate dark matter and/or inflation.
 The concrete  choice of the geometric
 Lagrangian determines further details of the theory.
 The most natural geometric models
 look similar to recently proposed brane models
 of cosmology usually derived from string theory.

\end{abstract}

 The history of science teaches us not to completely forget beautiful
 and logically consistent papers of the past that were not understood
 in time. Even though not recognized by the contemporaries
 (and, often, by the authors) some of them happen to become of interest
 many years after their publishing. One can recall sufficiently many
 examples of such work and here we discuss a misunderstood and forgotten
 model based on work of three eminent scientists (Weyl, Eddington and
 Einstein) that was reinterpreted and generalized
 in \cite {ATF}-\cite{ATFs}.

 By the end of 1922, Einstein deeply studied and seriously reconsidered
 attempts of Weyl and Eddington (see \cite{Weyl} - \cite{Ed}) to construct
 an affine modification\footnote{In 1918, Weyl introduced a special
 symmetric non-Riemannian
 connection depending on a metric tensor and on a vector field
 (`Weyl's connection'), which he attempted  to identify with
 the electromagnetic potential. His theory was severely criticized and
 is mostly remembered because in it he first introduced a fairly general
 concept of the gauge symmetry. }
  of his general relativity. In 1923 he published
  three beautiful and concise papers \cite{Einstein1}
  later summarized in \cite{Einstein2} and soon forgotten (but see brief
 discussions  in \cite{Erwin}, \cite{Pauli}).
 The most clear exposition of Einstein's approach is given
 in \cite{Einstein2} while the most beautiful model was proposed
 in the first paper of the series \cite{Einstein1}.
 Here we only briefly summarize general principles, which can
 more or less naturally restrict possible choice of the
 physical models. Then a simple model satisfying these principles
 and generalizing Einstein's first paper, which we call the Einstein -
 Eddington model, will be introduced and compared to some
 recently discussed cosmologies  based on string theory.

 The most important properties of the affine theory are the
 following. {\bf 1.}~It predicts the existence of one or more
 vector fields with real or imaginary mass. {\bf 2.}~Its
 $D$-dimensional generalization predicts (after the simplest
 dimensional reduction) $(D-4)$ scalar fields with the same mass.
 {\bf 3.}~Both the vector and scalar fields couple to gravity only
 (being the part of the generalized gravitation). {\bf 4.}~The most
 natural effective (`physical') Lagrangian
 contains Eddington -  Einstein terms
 (nowadays often called Dirac - Born - Infeld terms).

 \emph{Einstein's key idea}
 was to derive the concrete form of the
 affine connection by applying the Hamilton principle to
 a generic Lagrangian depending on the generalized Ricci curvature.
 This assumption completely fixes a geometry,
 which does not coincide with Weyl's geometry,
 but belongs to the same simple class of connections
 introduced and discussed in \cite{ATF} - \cite{ATFs}.
 Einstein's unusual result was difficult to comprehend
 in the first half of the last century and it remains
 somewhat puzzling even these days.
 From the modern mathematics viewpoint, its origin could
 be ascribed to a sort of
        a mismatch between the  affine connection geometry
        and the Lagrangian `geometry'.
 At the moment, it is difficult to find a more detailed explanation.
 Possibly, this is an interesting mathematical problem.

 In Refs.\cite{ATF} - \cite{ATFs} we follow Einstein's approach
 and first construct a \emph{geometric Lagrangian density}
 having the dimension $L^{-D}$ (in units $c=1$).
 Then we show that, without a metric, one can use scalar densities
 of weight two constructed of pure geometric fields
 (see \cite{ATF} - \cite{ATFs}), the square roots of which give
 the desirable scalar densities of weight one. The effective
 \emph{physical Lagrangians} are derived from the geometric ones.
 A more detailed presentation of the main steps
 briefly discussed here can be found in \cite{ATF} - \cite{ATFs}.

 Here we first outline basic {\bf geometrical facts}
  and then concentrate on a physical model that looks most
 interesting for applications to cosmology.

 A general exposition of the theory of non-Riemannian spaces
 equipped with a \emph{symmetric connection} can be found in
 \cite{Eisen}, \cite{Erwin}, and in our previous papers.
  In general, the connection
 coefficients can be expressed in terms of a
 Riemannian connection $\Gamma^i_{jk}$ and of
 an arbitrary  third rank \emph{tensor} $a^i_{jk}$ that
 is symmetric in the lower indices
 \be
 \label{app1}
 \gamma_{jk}^i  \,=\, \Gamma_{jk}^i[g] + a^i_{jk} \,.
 \ee
 Here $g_{ij}$ is and arbitrary symmetric tensor and
 $\Gamma^i_{jk}[g]$ is its Christoffel symbol.
  More precisely, for any symmetric connection
  $\gamma_{jk}^i$, there exists
 a symmetric tensor $g_{ij}$ and a tensor
 $a^i_{jk} = a^i_{kj}$ such that (\ref{app1}) is satisfied.

 The curvature tensor $r^i_{jkl}$ can be defined
 in terms of $\gamma_{jk}^i$ by the standard general expression
 not using any metric,
 \be
 \label{1}
 r^i_{jkl} = -\gamma^i_{jk,l} + \gamma^i_{mk} \gamma^m_{jl}
 + \gamma^i_{jl,k} - \gamma^i_{ml} \gamma^m_{jk} \,.
 \ee
 Then, the Ricci-like (but \emph{non-symmetric}) curvature
 tensor can be defined by contracting the indices $i, l$:
 \be
 \label{2}
 r_{jk} \equiv r^i_{jki}
 = -\gamma^i_{jk,i} + \gamma^i_{mk} \gamma^i_{ji}
 + \gamma^i_{ji,k} - \gamma^i_{mi} \gamma^m_{jk}
 \ee
 (we again stress that $\gamma^i_{jk} = \gamma^i_{kj}$
 but $r_{jk} \neq r_{kj}$). Using only these tensors and the
 completely antisymmetric tensor density of the rank $D$,
 we can construct a quite rich geometric structure.

 The \emph{antisymmetric part} of the Ricci curvature
 $r_{ij}\,$ can be expressed in terms of the vector field
 $\gamma_i \equiv \gamma^m_{im}$
 or in terms of $a_i \equiv a^m_{im}~$, which
 differ by the gradient term $\p_i \ln \sqrt{|g|}$
 ($g \equiv \textrm{det}(g_{ij})$):
 \be
 \label{4}
 a_{ij} \equiv \half (r_{ij} - r_{j\,i})
 \equiv -\half (a_{i,j} - a_{j,i})
 \equiv  -\half (\gamma_{i,j} - \gamma_{j,i})\,.
 \ee
 We call this field \emph{vecton} and will see that it can be massive or
 tachyonic depending on a choice of the connection.
 This definition of the vecton is independent of the
 division of the connection (\ref{app1}) into the metric and
 non-metric parts.
 By the way, $r^m_{mij} = 2a_{ij}$.

 Introducing the covariant derivative $\nabla^{\gamma}_i$
  (with respect to the connection $\gamma$) we can write
  the symmetric part of the curvature as
 \be
 \label{4b}
 s_{ij} \equiv \half (r_{ij} + r_{j\,i})
 =  - \nabla^{\gamma}_m \gamma^m_{ij} +
 \half (\nabla^{\gamma}_i \gamma_j + \nabla^{\gamma}_j \gamma_i) -
 \gamma^m_{ni} \gamma^n_{mj} + \gamma^n_{ij} \gamma_n \,.
 \ee
 Using the `metric' covariant derivative
 $\nabla^g_i \equiv \nabla_i$ we can rewrite $s_{ij}$
 in the form
 \be
 \label{app3}
 s_{ij} = R_{ij}[g] - \nabla_m a^m_{ij} +
 \half (\nabla_i \, a_j + \nabla_j \, a_i) +
    a^m_{ni} a^n_{mj} - a^m_{ij} a_m \,,
  \ee
 where $R_{ij}[g]$ is the standard Ricci tensor of
 a Riemannian space with the metric $g_{ij}$.

 Now, suppose that
 $\textbf{c}^{ij} \equiv \sqrt{-g} \, c_{ij}$
 is an arbitrary tensor density.
 Then, its covariant derivative with respect to connection
 (\ref{app1}) is defined by
 \be
 \label{10c}
 \nabla^{\gamma}_i \, \textbf{c}^{kl} \,= \,\,
 \p_i \, \textbf{c}^{kl} + \,\gamma^k_{im} \, \textbf{c}^{ml} +\,
 \gamma^l_{im} \, \textbf{c}^{km}   -
 \gamma^m_{im} \, \textbf{c}^{kl} \, .
 \ee
 For any \emph{antisymmetric} density,
 $\textbf{c}^{ij} \equiv \textbf{f}^{ij} =  -\textbf{f}^{ji}$,
 it follows that
 \be
 \label{app7}
 \nabla^{\gamma}_i \, \textbf{f}^{ik} \,= \,
 \nabla^g_i \, \textbf{f}^{ik} \,= \,
 \p_i \textbf{f}^{ik} \,.
 \ee
 The \emph{symmetric} tensor density
  $\textbf{g}^{ij} \equiv \sqrt{-g}\, g^{ij}$
 obviously satisfies the equations
 \be
 \label{app8}
 \nabla^{\gamma}_i \, \textbf{g}^{ik} \,= \,
 \sqrt{-g}\, a^k_{im}\, g^{im} \,, \qquad
 \nabla^g_i \, \textbf{g}^{ik} \,= \, 0.
 \ee
 Eqs.(\ref{4b}) - (\ref{app8}) will be used in what follows.

 For a general symmetric connection one can introduce
 the concept of the geodesic curve (path),
 the tangent vector to
 which is parallel to itself at every point of the curve.
  The equations for the geodesic curves of any
  symmetric connection $\gamma^i_{jk}$
  can be written as
 \be
 \label{4c}
 \ddx^{\,i} +\, \gamma^i_{jk} \, \dx^j \, \dx^k = 0 \,,
 \ee
 where the dot denotes differentiating with respect to
 the so called `affine' parameter $\tau$ of the curve
 $x^i(\tau)$. Using the affine parameter we can compare
 the distances between points on the same curve.

 For a particular geodesics, the affine parameter
 is unique up to an affine transformation
 $\tau \mapsto \tau^{\prime} = a \tau + b$.
 Each connection define the unique set of paths, but
 all symmetric connections
 (with an arbitrary vector $\ha_k$)
 \be
 \label{4d}
 \hga^i_{jk} = {\gamma}^i_{jk} +
 \delta^i_j \, \ha_k + \delta^i_k \, \ha_j  \,,
 \ee
  define the same
 paths. The Weyl (conformal) tensor $W^i_{jkl}$
 of connection (\ref{4d}) is independent of $\ha_k$ while
 the Ricci tensor and its symmetric and antisymmetric
 parts are $\ha_i$-dependent
 (see \cite{Eisen} for more details).

 Therefore, an interesting class of connections is
 \be
 \label{4e}
 \hga^i_{jk} = {\Gamma}^i_{jk} [g] +
 \delta^i_j \, \ha_k + \delta^i_k \, \ha_j  \,,
 \ee
 where ${\Gamma}^i_{jk}[g]$ is a Riemannian connection
 (the Christoffel  symbol of a symmetric tensor $g_{ij}$).
 The paths of the connection $\hga^i_{jk}$ coincide with
 the geodesics of ${\Gamma}^i_{jk}[g]$, but the Ricci tensor
 of $\hat{\gamma}$ is symmetric \emph{if and only if}
 $\ha_i = \p_i \, \ha$
 with some scalar $\ha$. We see that connection (\ref{4e}) is
 maximally close to the Riemannian connection
 ${\Gamma}^i_{jk}[g]$
 and may be called a \emph{geodesically Riemannian}
 (`$g$-Riemannian') connection.
 Weyl and Einstein studied more general connections that belong
 to the following class introduced in \cite{ATF}, \cite{ATFn}:
 \be
 \label{a3}
 \gamma^i_{jk} = \Gamma_{jk}^i [g] +
  \alpha ( \delta^i_j \, \ha_k +  \delta^i_k \, \ha_j) -
   (\alpha - 2\beta) g_{jk}\, \ha^i  \,,
 \ee
  where  $\ha^i = g^{im} \ha_m$.
 The Weyl connection corresponds to $\beta = 0$ and
 the $g$-Riemannian connection, to  $\alpha = 2\beta$.
  Einstein derived the connection for the
  space-time dimension $D=4$, his result is
  $\alpha = -\beta = \sixth$ (it was generalized to any
  dimension in \cite{ATFs}).

 Using (\ref{app3}) it is easy to calculate the physically
 important expression for the symmetric part of the Ricci curvature.
 The terms linear in $A$ are equal to
 \be
 \label{app4}
 (\alpha + \beta) (\nabla_i \ha_j + \nabla_j \ha_i) +
 (\alpha - 2\beta)\, g_{ij} \nabla_m \ha^m  \,,
 \ee
 and the quadratic terms are
 \be
 \label{app5}
 \ha_i \ha_j \,\bigl[(\alpha - 2\beta)^2 -3\alpha^2 \bigl]
  \,+\,
 2 \,g_{ij} \ha^2 (\alpha - 2\beta) (\alpha + \beta) \,.
 \ee
 As we shall soon see the presence of the $\ha_i \ha_k$
 term in the expression for $s_{ij}$ signals that the vector
 field $a_i$ has in general a nonzero mass and that
 the sign of the first term in (\ref{app5})
 can be positive or negative (the second term in (\ref{app5})
 and  the linear terms in (\ref{app3}) in general
 do not vanish).
 In particular, for the Weyl, Einstein and g-Riemannian connections
 the quadratic terms are, respectively:
 \be
 \label{app6}
 \textrm{W:} \,\,\, -\half [\ha_i \ha_k - 2 g_{ik} \ha^2] \,,
 \qquad  \textrm{E:} \,\,\, \sixth \ha_i \ha_k \,, \qquad
  \textrm{g-R:} \,\,\, -{3 \over 4} \ha_i \ha_k \,.
 \ee

 Before we leave pure mathematics and turn to more physical
 problems, we should mention one of
 the characteristic properties  of
 symmetric connections. For applications
 of geometry to gravity, it is very important that at every
 point of the affine-connected space-time manifold there must exist
 a geodesic coordinate system, such that the connection coefficients
 are zero at this point. Using the above formulas it is easy to
 prove that such a coordinate system exists if and only if the
 connection is symmetric. For symmetric connections, the Fermi
 theorem about the existence of geodesic coordinates along
 the curves also holds
 (for the precise definitions and proofs see \cite{Eisen}).

 Let us turn to {\bf dynamics}. Weyl's approach
 to constructing a physical theory based on the affine geometry
 is direct (if we discard his ideas on `linear metric' and
 on lengths calibrating): he first chooses a particular geometry
 (\ref{a3}) with $\beta = 0$ and then constructs tensor equations
 that should generalize the Einstein equations. He tries to identify
 the antisymmetric part of the curvature with the electromagnetic
 field tensor but the $\ha_i \ha_k$ terms spoil this interpretation.
 He eventually guessed a Lagrangian which is similar to the Einstein
 theory coupled to a vector field with the mass term that cannot be
 removed and with the cosmological constant that he introduces
 `by hand'. Eddington tried to find a
 generalization of Einstein's theory by considering the most
 general nonsymmetric affine connection. In a discussion of possible
 scalar densities he suggests the simplest one
 (we call it \emph{Eddington's scalar density}),
  \be
 \label{3}
 {\Lag} \equiv \sqrt{ -\det(r_{ij})} \,
 \equiv \, \sqrt{ -r} \,,
  \ee
 This resembles the fundamental scalar density of the Riemannian
 geometry, $\sqrt{-\textrm{det}(g_{ij})} \equiv \sqrt{-g}$,
 and Eddington tried to directly identify $s_{ij}$ with the
 metric. If we in addition identify
 $a_{ij}$ with the electromagnetic field we get
 a Born-Infeld Lagrangian. However, Eddington did not succeed
 in constructing consistent equations.

 A consistent \emph{Lagrangian formulation}
  of the generalized theory was found by Einstein.
 His approach is conceptually different both from Weyl's and
 Eddington's ones  and consists of two stages.
   In the first  stage, he assumed
   that the general symmetric connection should be
  restricted by the Hamilton principle for a general Lagrangian
 density depending either on $r_{ij}$ (see the second
 paper\footnote{
 In the first paper Einstein uses as the Lagrangian the Eddington
 density but later he realized that in the first stage
 it is sufficient to suppose that the Lagrangian is an arbitrary
 scalar density depending on $r_{ij}$}.)
 or on $s_{ij}$ and $a_{ij}$ separately  (in the third paper).
 He gave no motivation for this assumption, but it is easy to see
 that the resulting theory in the limit $a_{ij}=0$  is
 consistent with the standard general relativity supplemented
 with a cosmological term. In this stage, Einstein succeeded in
 deriving the remarkable expression  for the connection
 (see (\ref{a3}) with $\alpha = -\beta = \sixth $)
 and the general expression for $s_{ij}$ depending on
 a massive (tachyonic) vector field and the metric
 tensor density $\textbf{g}^{ij}$.

 In the next stage, a concrete Lagrangian density
 ${\Lag}(s_{ij} \, , a_{ij})$  should be chosen.
 Einstein did not formulate any principle for selecting
 a Lagrangian, and both from geometric and physical standpoint
 his concrete choice seems sufficiently arbitrary,
 especially in the third paper where he essentially
 reproduced one of the Weyl results.
 We believe that his best choice was
 made in the first two papers and, indeed, very similar
 effective Lagrangians are considered in modern applications
 of the superstring theory to cosmology.
 We may try to formulate some properties of possible geometric
 Lagrangian densities ${\Lag}$ that are consistent with the
 Eddington-Einstein Lagrangian but allow for a more general
 class of them (with different mass terms, different dependence
 on $s_{ij}$, $a_{ij}$, $a_i$, in different space-time dimensions).

 Naturally, the Lagrangian must depend on
 \emph{tensor variables having a direct geometric meaning}.
 It is desirable that in the next stage they will acquire
 a \emph{natural physical interpretation}. As soon as we do not wish to
 fix the division of the connection into the metric and
 non-metric parts by Eq.(\ref{app1}), we can take
 the vector $\gamma_i$ (not $a_i$!),
 second-rank tensors,
 $s_{ij}$, $a_{ij}$,  $\gamma_{ij} \equiv \gamma_i \gamma_j$
 (if we used representation (\ref{app1}) for $\gamma^k_{ij}$
 we could add to this list
 $\bar{\gamma}_{ij} \equiv \gamma_k a^k_{ij}$, but this
 tensor implicitly depends on the metric and we must not
 use it at the first stage).  We can construct
 higher-rank tensors, but the tensors of the second rank
 (especially, the first three) look more fundamental
 from the physics point of view.

 Consider the \emph{second-rank tensors} as building blocks
 of the `geometric' Lagrangian. They all have
 the dimension $L^{-2}$
 (in the units $c=1$) and we can use as Lagrangian densities
 some homogeneous functions of the degree $D/2$ and
 dimension $L^{-D}$ that are independent
 on any dimensional constants. After integrating
 over $D$-dimensional
 volume element $dx^0 \bigwedge ... \bigwedge dx^{D-1}$
 we then get a dimensionless quantity playing a role of
 a \emph{geometric action}. The simplest Lagrangian density
 then depends on three second-rank tensors,
   \be
 \label{3a}
 {\Lag} = {\Lag} (s_{ij} , a_{ij}, \gamma_{ij}) \,,
 \ee
 and a density having  the correct dimension $L^{-D}$
 can easily be written:
 \be
 \label{3aa}
 {\Lag}_g = \sqrt{-\det(s_{ij} + \nu a_{ij} +
 \nu_1 \gamma_{ij})}\,.
 \ee
 Here we take the minus sign because $\det(s_{ij})<0$
 (due to the local Lorentz invariance) and we
 naturally assume that the same is true for
 $\det(s_{ij} + \nu a_{ij} + \nu_1 \gamma_{ij})$
 (to reproduce Einstein's general relativity in the limit
 $\nu, \nu_1 \rightarrow 0$). The $\nu$-parameters
 are dimensionless, we mainly introduce them to
 disentangle the scale of the mass parameter
 of the vector field from the cosmological constant.
 If we take the original Eddington - Einstein Lagrangian
 (\ref{3}), the mass squared will be of the order
 of the cosmological constant $\Lambda$ (see \cite{ATFn}).
 Lagrangian (\ref{3aa}) with $\nu_1 =0$ was proposed and
 studied in some detail in \cite{ATFn}.
 The general Lagrangian (\ref{3aa}) was first considered in
 Ref.\cite{ATFs}, where we also discussed a more general construction
 that allows to write other Lagrangians having the desired
 properties. Unfortunately, these generalized Lagrangians
 are more complicated both technically and conceptually,
 and we do not discuss them here.

 We emphasize that the Lagrangians (\ref{3}) and (\ref{3aa})
 are written in the form independent of $D$,  although
 the analytic expressions for the dependence of the
 determinants on  $s_{ij}$ and $a_{ij}$ essentially
 depend on $D$. Accordingly, the physical equations
 depend on the space-time dimension as we will shortly demonstrate.

 The starting point for Einstein (in his first paper
   of the series  \cite{Einstein1})
 was the action principle with
 Lagrangian density (\ref{3}) depending on 40 connection functions
 $\gamma^i_{kl}$.
 Varying the action with respect to these functions,
 he derived 40 equations that allowed him to find the
 expression for $\gamma^i_{kl}$ given by (\ref{a3}) with
 $\alpha = - \beta = \sixth \,$ (in the four-dimensional
 space-time).

 The main steps of his proof were reproduced in \cite{ATFn}.
 Here we somewhat generalize the derivation
 to an arbitrary dimension $D$ and assume that
 the geometric Lagrangian depends also on
 $\gamma_i \equiv \gamma^m_{im}$.
 We define the new tensor densities\footnote{
 Following Eddington's notation, we let boldface Latin letters
 denote tensor densities.
 The derivatives in (\ref{10}) and (\ref{10a})
 must be properly symmetrized,
 which is easy in concrete calculations.
  We tacitly assume that geometry
  has only a single dimensional constant,
 e.g., the cosmological constant $\Lambda$ with the dimension
 $L^{-2}$.  To restore the correct dimension in (\ref{10}) and
 (\ref{10a}),  we must then multiply the densities
  by $\Lambda^{(D-2)/2}$.}.
   \be
 \label{10}
  {{\p {\Lag}} \over {\p s_{ij}}} \equiv \textbf{g}^{ij} \,,
  \qquad
 {{\p {\Lag}} \over {\p a_{ij}}} \equiv \textbf{f}^{ij} \,,
  \qquad
 {{\p {\Lag}} \over {\p \gamma_i}} \equiv \textbf{b}^i \,,
  \ee
  and introduce a conjugate Lagrangian density
  ${\Lag}^* \, = \, {\Lag}^* (\textbf{g}^{ij} ,
  \textbf{f}^{ij} ,\textbf{b}^i)$
  by a Legendre  transformation,
 \be
 \label{10a}
 s_{ij} = {{\p {\Lag}^*} \over {\p \textbf{g}^{ij}}} \,, \qquad
 a_{ij} = {{\p {\Lag}^*} \over {\p \textbf{f}^{ij}}} \,, \qquad
 \gamma_i = {{\p {\Lag}^*} \over {\p \textbf{b}^i}} \,.
   \ee

   By varying ${\Lag}$ in $\gamma^i_{kl}$ and using
   the above definitions, we can then show that the conditions
   $\delta {\Lag} / \delta \gamma^i_{kl} = 0$ are equivalent
   to the following 40 equations
   \be
   \label{10b}
  2 \nabla^{\gamma}_i \, \textbf{g}^{kl} \,= \,\,
  \delta^l_i \, [\nabla^{\gamma}_m \, (\textbf{g}^{km} +
  \textbf{f}^{km}) - \textbf{b}^k ] +
 \delta^k_i \, [\nabla^{\gamma}_m \, (\textbf{g}^{lm} +
 \textbf{f}^{lm}) - \textbf{b}^l ] \,,
   \ee
 where $\nabla^{\gamma}_i$ is the covariant derivative with respect to the
 affine connection $\gamma^i_{jk}$. Remembering (\ref{app7})
 we define the vector density $\hat\textbf{a}^k$ by
 \be
 \label{10d}
 \p_i \, \textbf{f}^{ki} \,-\, \textbf{b}^k
 \, \equiv \, \hba^k ,
  \ee
  and then easily find that
 \be
   \label{10e}
  \nabla^{\gamma}_i \, \textbf{g}^{ik} \,= \,\,
  - {\frac{D+1}{D-1}} \hba^k  \,,
   \ee
 Now it is easy to find the equations from which
 the connection coefficients can be derived
 (as in the Riemannian case):
 \be
   \label{10f}
  \nabla^{\gamma}_i \, \textbf{g}^{kl} \,= \,\,
  - {\frac{1}{D-1}} \bigl(\delta_i^k \hba^l +
  \delta_i^l \hba^k \bigr).
   \ee
 Defining the Riemann metric tensor
 $g_{ij}$ by the equations
 \be
 \label{11}
 g^{kl} \sqrt{-g} \, = \, \textbf{g}^{kl} \,, \quad
 g_{kl} \, g^{lm} \, = \, \delta^m_k \,,
\ee
 we can then define the corresponding Riemannian
 covariant derivative $\nabla_i$, for which\,
 \be
 \label{11a}
 \nabla_i \, g_{kl} = 0, \quad
 \nabla_i \, g^{kl} = 0 .
  \ee

 Taking the above into account, we can now use (\ref{10f})
 to derive the expression  for $\gamma_{jk}^i$
 in terms of the metric tensor $g_{ij}$ and
 of  the vector $\ha^k \equiv \hba^k/\sqrt{-g}$,
 \be
 \gamma_{jk}^i  \,=\, \Gamma_{jk}^i[g] \,+\,
 \alpha_D \, \bigl[\, \delta^i_j \, \ha_k +
 \delta_k^i \, \ha_j - (D-1)\, g_{jk} \, \ha^i \bigr] \,,
 \label{a31}
 \ee
 which corresponds to $\alpha = \alpha_D$  and
  $\beta = \beta_D$ in (\ref{a3}), with
 \be
 \label{a32}
 \alpha_D \equiv [(D-1)(D-2)]^{-1} \,,
 \qquad
 \beta_D \equiv - [2(D-1)]^{-1}\,.
 \ee
  For $D=4$, this coincides with Einstein's result
 for the connection.
 If we add $\gamma_{ij}$ as an independent variable,
 the connection remains the same. Note also that the
 added variables remain non-dynamical and attempting
  to make them dynamical in the second stage `by hand'
 destroys the beauty of the original Einstein construction.

  We cannot go deeper into discussions of further relations
 between geometry of affine connections and dynamical
 principles. But the above results show that these
 relations are rather complex and we do not yet
 understand their nature.
 Indeed, we tried to add new natural variables
 into the geometric Lagrangian, but the
 class of connections obtained as an output of
 Einstein's approach did not change at all.
 It can be argued that there are many other, not yet
 explored options, but in reality, we do not even know
 how to obtain Weyl's or $g$-Riemannian connections
 following Einstein's approach.

 One of the possibilities could be to abandon some of
 Einstein's assumptions. The most serious drawback
 (or virtue, depending on a viewpoint) of his approach
 is that two  pairs of the basic variables of the theory,
 ($s_{ij}$,  $\textbf{g}^{ij}$) and
 ($a_{ij}$,  $\textbf{f}^{ij}$),
 having very different geometrical and physical
 meaning, are treated symmetrically. Definition
 (\ref{10}) looks quite natural for the metric
 density because Einstein's Lagrangian for the pure
 gravity theory is simply $\textbf{g}^{ij} R_{ij}$.
 But Einstein's definition of $\textbf{f}^{ij}$
 tacitly (and, as we see, wisely!)
 assumes that the geometric Lagrangian is
 independent of $\gamma_i$ or  $\gamma_{ij}$.
 This may look rather paradoxical, but, as we have seen,
 the mass term is dictated by the geometry because its
 germ, the term $\sim a_i a_j$, is already present in the
 expression for $s_{ij}$.\footnote{
 Therefore, it would be more natural to identify
 the field tensor of the massive vector field directly
 with $a_{ij}$, up to a necessary dimensional multiplier.}
 Its interpretation as the physical mass comes
 when we write an effective physical Lagrangian.
 Then the geometric Lagrangian generates only the kinetic terms
 and is in fact the Lagrangian of a brane.

 There are many other questions, which should be
 carefully discussed, but we postpone the discussion
 to future publications. Here, we present a simple example
 demonstrating how to eventually pass
 {\bf from geometry to physics}
 and to demonstrate a relation of the Einstein
 approach to the present-day concerns. Our discussion suggests
 that the geometric Lagrangian (\ref{3aa})
 with $\nu_1 =0$ is better motivated by geometry and physics
 than other ones. This Lagrangian is most natural and gives
 the effective physical Lagrangian belonging to a class
  widely discussed in relation to modern problems of cosmology.
 We only briefly describe this model which is, possibly, the
 simplest generalization of Einstein's general relativity.

  Pure geometry gives us equations (\ref{4})
  and (\ref{app3}). With $a^i_{jk}$ given by
 (\ref{a3}), their right-hand sides are given by
 $(a_{i,j} - a_{j,i})/2$, where
 $a_i = (D\alpha +2\beta) \,\ha_i$, and by the sum
 of $R_{ij}$ with expressions (\ref{app4}), (\ref{app5}).
 To derive $s_{ij}$ and $a_{ij}$ in terms of the
 `physical' variables $g_{ij}$ and $f_{ij}$ we
 must choose a Lagrangian (e.g., (\ref{3aa})) and
 then solve equations (\ref{10}) with respect to
 the geometric variables $s_{ij}$ and $a_{ij}$.
 Alternatively, if we know the conjugate Lagrangian
 ${\Lag}^* (\textbf{g}^{ij} ,\textbf{f}^{ij})$,
 we can directly calculate them using (\ref{10a}).

 In \cite{ATFn}, we reproduced Einstein's result
 of Refs.\cite{Einstein1}, \cite{Ed1} (in which it was not
 written explicitly but could easily be derived):
 \be
 \label{11e}
 {\Lag} \equiv \sqrt{ -\det(r_{ij})} \,=\,
 4\sqrt{-\det(\textbf{g}^{ij} + \textbf{f}^{ij})}
 \, \equiv \, 4\sqrt{-\det(g_{ij} + f_{ij})} \,
 =\, {\Lag}^* \,.
  \ee
 We emphasize that these equations are valid
 only in the four-dimensional theory.
 Note that the equality ${\Lag}^* = {\Lag}$ simply follows
 from the fact that ${\Lag}$ is a homogeneous function of
 the degree two but, in general,
  the concrete expression for ${\Lag}^*$
 must be obtained by a direct calculation.
 Now we can show that the relation like (\ref{11e})
 holds also for Lagrangian (\ref{3aa}) with $\nu_1 = 0$
 and $\nu \neq 0$, which we rewrite as
 \be
 \label{3b}
 {\Lag}_{\nu} \equiv \sqrt{ -\det(s_{ij} + \nu a_{ij})} \,.
  \ee
 This can be done by a direct computation but
 it is simpler to first dimensionally reduce ${\Lag}_{\nu}$.

 Consider the $D=4$ case and
  define a \emph{`spherical reduction'} not using
 any metric. Suppose that $s_{ij}$ and $a_{ij}$
 are functions of $(x^0,x^1)$ and that $a_2 = a_3 =0$
 (therefore, only $a_{01} = -a_{10} \neq 0$).
 We then assume that the symmetric matrix has
 the following nonzero elements:
 $s_{ij} = \delta_{ij}\, s_i \,$,
 $\,s_{01} = s_{10}$ (our result will
 not change if also $s_{23} \neq 0$).
 By explicitly deriving $s_{ij} + \nu a_{ij}$, we can
 find $\textbf{g}^{ij}$ and $\textbf{f}^{ij}$
    (using (\ref{10})) and hence express
 $\det(\textbf{g}^{ij} + \lambda \textbf{f}^{ij})$
  in terms of $s_{ij}$ and $a_{ij}\,$:
 \be
 \label{11f}
 16 \det(\textbf{g}^{ij} + \lambda \textbf{f}^{ij}) =
 \det[s_{ij} + (\nu^2 \lambda) \,a_{ij}] .
 \ee
 It follows that choosing $\lambda = \nu^{-1}$ we
 have\footnote{
 We ignore the dimensional constants while
 working mainly with geometrical theory, where
 presumably exists just one dimensional constant
 $\Lambda$ (with $c=1$). Then  emergence of
  some dimensionless parameters may signal
  that there exist other dimensional
 constants (e.g., different scales in symmetric and
 antisymmetric sectors of geometry may be described
 by introducing our parameter $\nu$). We eventually
 restore dimensions in the effective physical Lagrangian.}
 \be
 \label{11g}
 {\Lag}_{\nu} = \sqrt{-\det(s_{ij} + \lambda^{-1} \,a_{ij})} =
  4 \sqrt{-\det(\textbf{g}^{ij} +
  \lambda \textbf{f}^{ij})}  =
  4 \sqrt{-\det(g_{ij} +
  \lambda f_{ij})}
  = {\Lag}^*_{\lambda} \,,
 \ee
 where the sign and normalization are arbitrary chosen in relation
 to the cosmological interpretation.
 This result is written in the form not implying the
 spherical reduction, and we suppose it is true in a general
 four-dimensional theory. In arbitrary dimension ($D \neq 2$)
 it must be somewhat modified as was first shown in \cite{ATFn}.

  To similarly treat the {\bf higher dimensional case}
  we first reduce the $D$-dimensional Lagrangian
 to the dimension four. For simplicity, let us consider $D = 5$.
 Then the field $a_k \, (k=0,..,4)$ depends only on
 $x_i \, (i=0,..,3)$, $a_{ij} =\half (\p_j a_i - \p_i a_j)$,
 and $a_{4i} = 1/2 \,(\p_i \,a_4)$.
 Therefore the terms containing $a_{4i}^2$
 should be interpreted in four dimensions as kinetic terms of
 the \emph{scalar field} $a_4$.\footnote{
 It can be seen that this scalar field is massive or tachyonic.
 In the simplest reduction, its mass coincides with
 that of the vecton.}
 Applying  spherical reduction to the resulting
 four-dimensional Lagrangian, we can construct
 a two-dimensional model effectively describing spherically
 symmetric solutions of the four-dimensional gravity
 coupled to the vecton and to the scalar fields.
 To get the corresponding Lagrangian we
 derive the determinant of the matrix
 \be
 s_{ij} + \nu a_{ij} \equiv s_i \de_{ij} +
 (\de_{0i} \de_{1j} + \de_{0j} \de_{1i})
 (s_{ij} + \nu a_{ij}) +
 (\de_{i4} + \de_{j4}) a_{ij} \,,
 \label{12}
 \ee
 where $a_{ij}$ are defined in terms of $a_i\,$, and
 all the functions in (\ref{12}) depend on $x^0, x^1$
 (thus $a_{24} = a_{34} = 0$). The determinant is
 \be
 \det(s_{ij} + \nu a_{ij}) = \prod_{i=0}^4 s_i
 \, [1 + \ts^2_{01} - \nu^2 ( \ta^2_{01} + \ta^2_{04}
 - \ta^2_{14})] ,
  \label{12a}
 \ee
 where we define
 $\tm_{ij} \equiv m_{ij} \, |s_i s_j|^{-1/2}$.
 The determinant obviously has zeroes and thus its
 square root is always singular. Therefore, the
 corresponding two-dimensional dilaton gravity
 describing spherically symmetric solutions
 is rather unusual and complex.
 By further reductions to static or cosmological
 configurations
 we can construct corresponding one-dimensional dynamical
 systems describing static states with horizons as well as
 cosmological models.
 The cosmological models look realistic enough because
 they incorporate a natural sources of the dark energy,
 inflation, and, possibly, some candidates for the dark
 matter (for a more detailed discussion see
 \cite{ATF}, \cite{ATFn}).

 Before presenting a simplest cosmological model, we
 write the general $D$-dimensional theory.
 In addition to predicting scalar fields, the higher-dimensional
 Lagrangians differ from the ones usually considered
 in modern brane cosmology. In fact, while the square-root
 Lagrangian ${\Lag}$ produces the square-root Lagrangian
  ${\Lag}^*$, which gives the so-called DBI-like term
 in the effective physical Lagrangian
 (see many examples in \cite{Born} - \cite{Langlois}),
 our higher-dimensional Lagrangian essentially
 depends on $D$:
 \be
  \sqrt{-\det(s_{ij} + \nu a_{ij})} =
   [-2^D \det(\textbf{g}^{ij} +
  \lambda \textbf{f}^{ij})]^{1/(D-2)} =
  \sqrt{-g} \ [- 2^D  \det (\delta_i^j +
 \lambda f_i^j)]^{1/(D-2)} \,,
 \label{dr}
 \ee
 which coincides with (\ref{11g}) for $D=4$.
  Following \cite{ATFn}, we may write the
 corresponding {\bf physical Lagrangian}
  \be
 \label{8}
 \Lag_{eff} =  \sqrt{-g} \,
 \biggl[ -2 \Lambda \,[\det(\delta_i^j +
 \lambda f_i^j)]^{1/(D-2)} +  R(g) +
 c_a \, g^{ij} a_i a_j \biggr] \,,
 \ee
  which should be varied with respect to the metric and the
  vector field; $c_a$ is a parameter depending on $D$
  (Einstein's first model is obtained for $D=4$ and
  $c_a = 1/6$).   When the vecton field is zero,
  we have the standard Einstein gravity
  with the cosmological constant. Making the dimensional
  reduction from $D=5$ to $D=4$,
  we obtain the Lagrangian describing
 the vecton $a_i$, $f_{ij} \sim \p_i a_j - \p_j a_i$
 and $(D-4)$ scalar fields $a_k, \, k = 4,..,D$.

  The theory (\ref{8}) is very complex,
  even at the classical level. Its spherically symmetric
  sector is described by a (1+1)-dimensional dilaton gravity
  coupled to one massive vector and to several scalar fields.
  If the mass of the vector field is zero and the scalar
  fields vanish\footnote{Such models can be derived
  by dimensional reductions of some higher-dimensional gravity
  and supergravity theories, see, e.g,
  \cite{ATF4} - \cite{ATF3} and references therein.},
   the \emph{ dilaton gravity is classically integrable with
   a rather general dependence of the Lagrangian
 on the massless Abelian gauge fields},
 $X(\phi, F^2)$, where $F^2 = F_{ij} F^{ij}$ and $\phi$
 is the dilaton field, see \cite{ATF4}-\cite{ATF1}.
 If $\mu^2 \neq 0$,
 the theory is certainly not integrable even with vanishing
  scalar fields. It is also not easy to analytically construct
  its physically interesting approximate
  solutions\footnote{At first sight, a perturbation theory
  in $\mu^2$ seems to be a viable alternative to numeric
  approximations but, when $\mu^2 = 0$, an additional gauge
  symmetry emerges that makes it difficult to estimate
  the validity of the approximation, especially, in the
  physically important asymptotic regions.}.

  Further dimensional reductions
  to one-dimensional static or cosmological theories
  also give non-integrable dynamical systems although
  some approximate solutions can possibly be derived.
  The naive cosmological reduction of four-dimensional
  theory (\ref{8}) can be written using the metric
  \be
 ds_4^2 = e^{2\alpha} dr^2 + e^{2\beta}
 d\Omega^2 (\theta , \phi) - e^{2\gamma} dt^2  \,,
 \label{eq1}
 \ee
 where $\alpha, \beta, \gamma$ depend on $t$
  and $d\Omega^2$ is the metric on the
  two-dimensional sphere.\footnote{
  The function $\beta(t)$
  is the two-dimensional dilaton field and,
  usually, it is supposed
  that $\alpha = \beta$ (isotropy condition).
  With the massive vector field $A_i(t)$,
  this is not possible because the equations
  of motion require $A_0 \equiv A_t=0$ and
  $A_1 \equiv A_r \neq 0$,
  which obviously gives an anisotropic configuration,
  see \cite{ATF}, \cite{ATFn}.}
  Now the effective cosmological
  (one-dimensional) Lagrangian corresponding
  to theory (\ref{8}) in the $D=4$ case is
 \be
 \cL_c =  -2 e^{2\beta} \biggl[e^{\alpha - \gamma}
 (\dbe^2 + 2\dbe \dal) +
 \Lambda \sqrt{e^{2(\alpha + \gamma)} - \lambda^2 \dA^2} +
  \half \, \mu^2 A^2 e^{-\alpha + \gamma}  \biggr] \,.
 \label{eq10}
 \ee
 As $\gamma$ is obviously a Lagrange multiplier we
 can fix the remaining gauge freedom by choosing
 $\gamma = -\alpha$.\footnote{ The standard gauge fixings are
 $\gamma = 0$  or $\gamma = \alpha$; in \cite{ATFn}
 we also used the gauge fixing
 $\gamma = 3\rho \equiv \alpha + 2 \beta$.
 Varying the Lagrangian multiplier $\gamma$ gives the
 \emph{energy constraint}, i.e. vanishing of the Hamiltonian.}
 Using this gauge and denoting the anisotropy function by
 $3\sigma \equiv \beta - \alpha$, we have the gauge fixed
 Lagrangian
  \be
 \cL_c =  -2 e^{2\beta} \biggl[3 e^{2\alpha}
 (\drho^2 - \dsig^2) +
 \Lambda \sqrt{1 - \lambda^2 \dA^2} +
  \half \, \mu^2 A^2 e^{-2\alpha}  \biggr] \,,
    \label{12b}
  \ee
 where $\alpha = \rho - 2\sigma$ and
 $\gamma = \rho + \sigma$.

 Up to the dilaton multiplier $e^{2\beta}$, the second
 term in (\ref{12b}) is the DBI (or, 0-brane) Lagrangian.
 If we consider constant metric functions $\alpha, \beta$,
 and denote $M_A \equiv 2\lambda^2 \Lambda e^{2\beta}$, we
 see that the 0-brane term is the relativistic Lagrangian
 of a particle with the mass $M_A$ (the analog of
 the velocity of light $c$ is
 $\lambda^{-1} \equiv \bar{c}$).
  Introducing the canonical
 momenta $p_{\rho}, p_{\sigma}, p_A$ we find the Hamiltonian
 (one should not forget that $M_A$ depends on $\beta(t)$):
 \be
 {\cal H} = \bar{c} \sqrt{p_A^2 + M_A^2 \, \bar{c}^2} +
 \mu^2 A^2 e^{2(\beta-\alpha)} +
 {1\over 24} \, e^{2(\beta + \alpha)} \,
 (p^2_{\sigma} - p^2_{\rho}) \,=\,0 .
 \label{12c}
 \ee
 If $\mu^2 = 0$, the momentum $P_A$ is the integral of motion
 and we get an  integrable 1-dimensional dilaton gravity.
 (with $\mu^2 \neq 0$, it is not integrable and rather
 unusual theory). If $\alpha$ and $\beta$ vary much slower than
 $A(t)$ this is a more tractable model of a relativistic
 `particle' with the slowly varying time dependent mass
 $M_A$ in a simple potential having time dependent parameters.
 A simpler effective `particle' model was used by Gribov
 for discovering the famous \emph{Gribov copies}.
 One may hope that a similar interpretation of the theory
 (\ref{12c}) will help to understand some unusual qualitative
 features of our generalized gravity.

 For small $A$ and slowly varying gravitational fields,
 one can also use the small-field approximation
 (see \cite{ATF}, \cite{ATFn}), which is
  formally equivalent to expanding (\ref{8}) in powers
  of $\lambda^2$. Keeping only the first-order
  correction we then obtain a nice-looking
  field theory:
 \be
 {\Lag}_{eff}\, \cong \,  \sqrt{-g}\,
 \biggr[R[g] - 2\Lambda -
 \kappa \biggr(\half F_{ij} F^{ij} + \mu^2 A_i A^i
 + g^{ij} \p_i \psi \, \p_j \psi + m^2 \psi^2 \biggl) \biggl]\,,
 \label{13a}
 \ee
 where $A_i \sim a_i$, $F_{ij} \sim f_{ij}$,
 $\kappa \equiv G/c^4$
 and we use the CGS dimensions. Note that here we choose
 the standard normalization of the fields and thus the
 dimensionless parameters of the theory ($D$, $\lambda$)
 are hidden in the masses $\mu$ and $m$. Note also that for
 Einstein's geometry the masses are imaginary, but we should
 study the general case when they may also be real.

 This simplified theory still keeps
 traces of its geometric origin: the simplest form of
 the dark energy (the cosmological constant $\Lambda$),
 massive (or tachyonic) vector and scalar fields, which
 can describe inflation and/or imitate  dark matter.
 The most popular inflationary models require a few
 massive scalar particles usually called inflatons
 (see, e.g.,  \cite{Linde} - \cite{Weinberg}).
 Without massive scalar fields, there is no
 simple inflation mechanism with one massive
 vecton. However, with the tachyonic vecton (see \cite{Ford})
 or  with several massive vector particles,
 it is probably easier to find more realistic inflation models
 (see  \cite{Bertolami} - \cite{Koivisto}; some of
 these papers also discuss a possible role of massive vector
 particles in dark energy and dark matter mechanisms).

 {\bf In conclusion}, we note that the geometrical and
 dynamical models discussed in this paper are not well understood,
 both conceptually and technically. Much work on them should
 be done before a realistic cosmological model could be
 constructed. In particular, one should study the relation
 between the geometry and dynamics discovered by Einstein.
 Possibly, one shall find behind it some symmetry principles
 which are not yet understood. One should also study
 more general theories. For example, why we not add to the
 geometric Lagrangian the terms quadratic in the
 curvature tensor that can be constructed not using any
 metric? Of course, the Eddington - Einstein Lagrangian
 and its simplest generalizations discussed here are most
 beautiful and are closely related to the modern theory of
 branes, but this is not a good enough argument for
  restricting alternative geometric proposals.
  The new part of the connection
 $a^i_{jk}$ is a tensor that can generate some higher spin
 fields and we must have some serious arguments for
 excluding this possibility from the very beginning.

 Finally, we must clearly state once more that the generalization
 of gravity considered here has nothing to do with other
 matter fields. It is not suggesting any unification of
 gravity with other forces of nature and with the standard matter.
 The true meaning of it and its unexpected relation to
 recent discoveries and ideas in cosmology is a real puzzle.
 Possibly, a  role of this theory is to replace
 the standard gravity inside the string theory
 which did not yet completely succeed in giving
  a simple and natural explanation of dark energy,
  inflation, and dark matter.

 \bigskip
 \bigskip

 It is a great sorrow to dedicate this article to the dear
 memory of Volodya Gribov and not to hear his sharp critical
 and highly stimulating remarks on its content.
  I realize that the ideas treated here might look
 to Volodya a bit far from physics he liked,
 but his incredible ability  to penetrate deep to the heart of
 any problem would certainly help to solve a puzzle
 left to us by three great scientists of the last century.

 \bigskip
 \bigskip

 {\bf Acknowledgment:}
  This work was supported in part by the Russian Foundation
 for Basic Research (Grant No. 09-02-12417 ofi-M).
 The author is also thankful to CERN-TH members and,
 especially, to L.Alvarez-Gaume for kind hospitality at
 CERN, where this paper was completed.


\newpage

\begin{thebibliography}{99}
 %
 \bibitem{ATF}A.T.~Filippov, `On Einstein - Weyl unified model of
 dark energy and dark matter', arXiv:0812.2616v2 [gr-qc] (2008).
 %
 \bibitem{ATFn}A.T.~Filippov, \emph{Theor. Math. Phys.}
 \textbf{163}(3) (2010) 753-767, \\
 arXiv:1003.0782v2 [hep-th] (2010).
 %
 \bibitem{ATFs}
 A.T.~Filippov, `Affine generalizations of gravity in the light
 of modern cosmology', arXiv:1008.2333v2 [gr-qc] (2010).
 %
 \bibitem{Weyl}H.~Weyl, \emph{Raum-Zeit-Materie}, Springer, Berlin, 1923
 (1-st ed. 1918; \\
 English translation 1950).
 %
 \bibitem{Ed1}A.S.~Eddington, \emph{Proc. Roy. Soc. London A},
 {\bf 99} (1919) 104-122.
 %
 \bibitem{Ed}A.S.~Eddington, \emph{The mathematical theory of relativity},
 Cambbridge Univ. Press, \\
 New York, 1923 (German translation of the 2-nd ed. 1925).
%
 \bibitem{Einstein1}A.~Einstein, \emph{Sitzungsber. Preuss.
 Akad. Wiss., Phys.-Math.}, (1923) 32-38, 76-77, \\
 137-140.
 %
 \bibitem{Einstein2} A.~Einstein, \emph{Nature},
 \textbf{112}, (1923) 448-449; `Eddington's Theorie und
 Hamiltonisches Prinzip', Appendix to the book:
 A.Eddington, \emph{Relativit\"{a}ts theorie in
 mathematischer Behandlung}, Springer, Berlin, 1925.
 %
 \bibitem{Erwin}E.~Schr\"{o}dinger, \emph{Space - time structure},
 Cambbridge Univ. Press, New York, 1950.
 %
 \bibitem{Pauli}W.~Pauli, `Relativit\"{a}tstheorie'
 in: \emph{Enzykl. d. Math. Wiss.}, Vol.5, Teubner, Leipzig (1921),
 539-775; ~~~
 \emph{Theory of Relativity}, Pergamon Press, Cambridge,
 New York, 1958.
 %
 \bibitem{Eisen}L.P.~Eisenhart, \emph{Nonriemannian geometry},
 Amer. Math. Soc. Publ. New York, 1927.
 %
 \bibitem{Born}M.~Born, \emph{Proc. Roy. Soc. london A},
 \textbf{143} (1933/34) 410-437; \\
 M.~Born and L.~Infeld, \emph{Proc. Roy. Soc. london A},
 \textbf{144} (1934) 425-451; \\
 \textbf{147} (1934) 522-546;
 \textbf{150} (1935) 141-166.
 %
  \bibitem{Dirac}P.A.M.~Dirac, \emph{Proc. Roy. Soc. London A},
 \textbf{268} (1960) 57-67.
 %
 \bibitem{Callan}C.~Callan and J.~Maldacena, `Brane dynamics from
from the Born - Infeld action', hep-th/9708147.
 %
 \bibitem{Gibbions}G.W.~Gibbons, `Born - Infeld particles and
 Dirichlet p-brane', hep-th/9709027.
 %
 \bibitem{Deser}S.~Deser and G.W.~Gibbons,
 \emph{Class. Quant. Grav.} \textbf{15} (1998) L35-L39; \\
 arXiv:hep-th/9803049v1 (1998).
 %
  \bibitem{Arkady}A.~Tseylin, `Born - Infeld action, supersymmetry and string
 theory', hep-th/9908105v5.
   %
 \bibitem{Townsend}P.K.~Townsend, `Brane theory solitons',
 hep-th/0004039.
 %
 \bibitem{Schwarz}J.~Schwarz, `Comments on Born - Infeld theory',
 hep-th/0103165.
 %
 \bibitem{Gibbons1}G.W.~Gibbons, `Aspects of Born - Infeld theory
 and string M-theory', hep-th/010659.
 %
 \bibitem{Banados}M.~Ba\~{n}ados,
 \emph{Phys. Rev. D}, \textbf{77} (2008) 123534;
  arXiv:0801.4103v4 [hep-th] (2008).
 %
 \bibitem{Langlois}D.~Langlois, S.~Renaux-Petel and D.A.~Steer,
 \emph{J. Cosmol. Astropart. Phys.}, \\
 \textbf{0904} (2009) 021;
  arXiv:0902.2941v1 [hep-th] (2009).
 %
 \bibitem{ATF4}A.T.~Filippov, \emph{Modern Phys. Lett. A},
 \textbf{11} (1996) 1691-1704; \\
 \emph{Internat. J. Mod. Phys. A},
 \textbf{12} (1997) 13-22.
 %
  \bibitem{VDA1}{V.~de Alfaro and A.T.~Filippov, `Integrable low
dimensional theories describing higher dimensional branes, black
holes, and cosmologies', hep-th/0307269.}
%
 \bibitem{ATF1}{V. de Alfaro and A.T. Filippov, `Integrable low
dimensional models for black holes and cosmologies from high
dimensional theories', hep-th/0504101.}
  %
 \bibitem{ATF3}{A.T. Filippov, `Some Unusual Dimensional Reductions of
Gravity: Geometric Potentials, Separation of Variables, and Static -
Cosmological Duality', hep-th/0605276.}
 %
 \bibitem{Linde}A.~Linde, `Particle physics and inflationary cosmology',
 arXiv:hep-th/0503203v1 (2005).
 %
 \bibitem{Starobin}V.~Sahni and A.~Starobinsky,
 \emph{Internat. J. Mod. Phys. D}, \textbf{15} (2006) 2105-2132; \\
 arXiv:astro-ph/0610026v3 (2006).
 %
 \bibitem{Mukhanov1}V.~Mukhanov, \emph{Physical foundations of cosmology},
 Cambridge Univ. Press, \\
 New York, 2005.
 %
 \bibitem{Rubakov}V.~Rubakov and D.~Gorbunov,
 \emph{Introduction into the theory of early Universe}, \\
  Vols. 1 and 2, Moscow, 2008-2009 [in Russian].
 %
 \bibitem{Weinberg}S.~Weinberg, \emph{Cosmology},
 Oxford Univ.Press, Oxford, 2008.
 %
 \bibitem{Ford}L.H.~Ford, \emph{Phys. Rev. D}, \textbf{40} (1989) 967-972.
 %
 \bibitem{Bertolami}M.C.~Bento, O.~Bertolami, P.V.~Moniz,
 J.M.~Mour\~{a}o,  and P.M.~S\'{a}, \emph{Class. Quant. Grav.},
   \textbf{10} (1993) 285-298;   arXiv:gr-qc/9302034v2 (1993).
 %
 \bibitem{Armen}C.~Armendariz-Pic\'{o}n, \emph{J. Cosmol. Astropart. Phys.},
  \textbf{0407} (2004) 007.
 %
 \bibitem{Mukhanov}A.~Golovnev, V.~Mukhanov and V.~Vanchurin,
 \emph{J. Cosmol. Astropart. Phys.}, \\
 \textbf{0806} (2008) 009.
 %
 \bibitem{Golovnev}A.~Golovnev and V.~Vanchurin,
  Phys. Rev. \textbf{D 79} (2009) 103524.
 %
 \bibitem{Koivisto}T.S.~Koivisto and D.F.~Mota, `Anisotropic
 dark energy: dynamics of backgroud and perturbations',
 ArXiv:o801.3776v.2 [astro-phys] (2008).

 %



 \end{thebibliography}
 \end{document}